# LinkedCT: A Linked Data Space for Clinical Trials

**Oktie Hassanzadeh, M.Sc.[1], Anastasios Kementsietsidis, Ph.D.[2], Lipyeow Lim, Ph.D.[2], Renée J. Miller, Ph.D.[1], Min Wang, Ph.D.[2]**
[1] University of Toronto, Ontario, Canada;
[2] IBM T.J. Watson Research Center, Hawthorne, NY, U.S.A.

**Abstract**

*The Linked Clinical Trials (LinkedCT) project aims at publishing the first open semantic web data source for clinical trials data. The database exposed by LinkedCT is generated by (1) transforming existing data sources of clinical trials into RDF, and (2) discovering semantic links between the records in the trials data and several other data sources. In this paper, we discuss several challenges involved in these two steps and present the methodology used in LinkedCT to overcome these challenges. Our approach for semantic link discovery involves using state-of-the-art approximate string matching techniques combined with ontology-based semantic matching of the records, all performed in a declarative and easy-to-use framework. We present an evaluation of the performance of our proposed techniques in several link discovery scenarios in LinkedCT.*

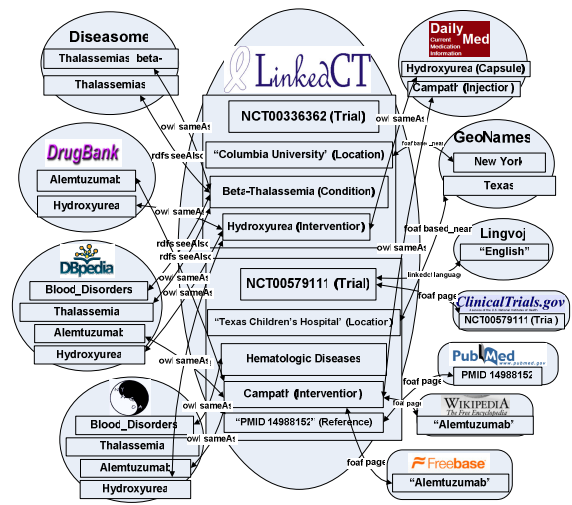

**Figure 1.** Sample interlinked entities in LinkedCT[*].

## Introduction

Clinical trials data are used extensively in drug development and clinical research[1]. The data collected in trials are widely used by a variety of different individuals for different purposes. Creating a single web data source of clinical trials data that is highly interlinked with existing medical data sources and that can be queried with semantically rich and complex queries, is a challenging task. This is the goal of the Linked Clinical Trials (LinkedCT) project. Such a web data source could significantly enhance the discovery of clinical trials in order to match patients with trials, perform advanced studies and enable *tailored therapeutics*[2].

The LinkedCT data space (`http://linkedct.org`) is published according to the principles of publishing *linked data*[3]. These principles greatly enhance adaptability and usability of data on the web, either by humans or machines. Each entity in LinkedCT is identified by a unique HTTP dereferenceable Uniform Resource Identifier (URI). When the URI is looked up, related RDF statements about the entity is returned in HTML or RDF/XML based on the user's browser. Moreover, a SPARQL endpoint is provided as the standard access method for RDF data.

Another important aspect of linked data publication is providing typed links to other relevant information sources for each resource. Without establishing such links, the online sources will resemble *islands of data* (or *data silos*), where each island maintains only part of the data necessary to satisfy the information needs of the particular users of that data source. In such a setting, apart from finding relevant data in a given island, users are also responsible for finding which other islands might hold relevant data, and what the specific pieces of relevant data are within these islands. To illustrate, consider the simple case of matching a patient with a clinical trial, where a clinician is trying to match a patient in "Westchester Medical Center" who is using the drug "Campath" with a related trial. For this, the clinician needs to identify the sources of information about trials, and in addition she first needs to identify all the locations close to where the patient lives, possibly using a source of geographical information (e.g., GeoNames[†]), and then find all the synonyms (generic names and brand names) of the drug using perhaps a drug data source (e.g., DrugBank[‡]). All of this is needed in order to identify, for example, a trial at "Columbia University" on "Alemtuzumab" (the generic name of the drug "Campath").

---
[*] Only the links to external entities are shown.
[†] `http://geonames.org`
[‡] `http://drugbank.ca`

Previous research in finding semantic links between data items has mainly focused on a more restricted version of the problem, namely, *entity resolution* or *record linkage*, i.e., the identification of entities in disparate sources that represent the same *real-world* entity[4]. Yet, the general problem investigated here considers links between entities that are not necessarily identical, although they are semantically related. For example, in Figure 1, "Columbia University" is linked with "New York" using a link of type `based_near` and "Beta-Thalassemia" is linked with "Blood Disorders" with link type `is_a_type_of`. In practice, the type of the links is usually specified using existing vocabularies, for example, `owl:sameAs` specifies that two entities are the same, `rdfs:seeAlso` specifies that the target resource may contain additional information about the source, and `foaf:based_near` specifies that the source entity is located in the target resource (which is of type location). Discovering such typed links is currently an area of active research.

**Triplification of Trials Data**

In this section, we outline the linked data generation (or *triplification*) of trials data in LinkedCT. The core of the LinkedCT dataset is derived from ClinicalTrials.gov[5], a registry of clinical trials conducted in the United States and around the world. As of November 2008, this registry contained information about more than 60,000 trials conducted in 158 countries. Each trial is associated with a brief description of the trial, related conditions and interventions, eligibility criteria, sponsors, locations, and several other pieces of information. The data on ClinicalTrials.gov is semi-structured and is available in HTML and XML formats. Transforming the data into RDF involves several steps. First, the XML collection of the trials needs to be crawled and stored locally. This results in a collection of over 60,000 XML files. Second, the XML data needs to be mapped into relational format. The mapping is obtained by manually inspecting the data and the Document Type Definition (DTD) of the data. In addition to all the entities and facts in the data, the mapping also extracts the relationships between the entities and their types. The relational schema needs to be reasonably normalized to ensure performance and avoid update anomalies when the data is refreshed. For LinkedCT, we have used a hybrid relational-XML DBMS (IBM DB2) to create relational views over the XML data, and to materialize the views to enhance the performance.

Third, the relational data is mapped into RDF format. We use the popular D2RQ tool for transforming the relational data to RDF[6] and the companion D2R server for publishing the relational data as a linked data source. Table 1 shows the overall statistics of the data and Table 2 shows the statistics of some of the entities in the data space.

**Table 1.** Overall Statistics.

| Total number of RDF triples | 7,011,000 |
|---|---|
| Number of Entities | 822,824 |
| Number of links to external linked data sources | 308,904 |

**Table 2.** Sample Entities.

| Entity | Count |
|---|---|
| Trials | 60,520 |
| Condition | 14,243 |
| Intervention | 67,271 |
| Location | 222,115 |
| Collaborator Agency | 7,071 |
| Overall Official | 52,660 |
| Primary Outcomes | 55,761 |
| Reference | 45,110 |
| Criteria | 73,688 |
| **Total** | **822,824** |

**Interlinking LinkedCT with Other Data Sources**

After the triplification of the trials data, we are faced with the challenge of finding useful links between the data in LinkedCT and other data sources. We have interlinked LinkedCT with several linked data sources of medical information: DBpedia[§] and YAGO (linked data sources derived from Wikipedia articles, containing many useful pieces of information about diseases and drugs), DailyMed[**] (published by the National Library of Medicine, providing high quality information about marketed drugs), Diseasome[††] (containing information about 4,300 disorders and disease genes linked by known disorder–gene associations for exploring known phenotype and disease gene associations and indicating the common genetic origin of many diseases), DrugBank (a repository of roughly 5,000

---

[§] `http://dbpedia.org`
[**] `http://www4.wiwiss.fu-berlin.de/dailymed/` (derived from `http://dailymed.nlm.nih.gov/`)
[††] `http://www4.wiwiss.fu-berlin.de/diseasome/` (derived from `http://www.ncbi.nlm.nih.gov/omim`)

FDA-approved drugs), Bio2RDF's PubMed (Bio2RDF[‡‡] project's linked data version of PubMed), GeoNames (linked data source of geographical information) and Lingvoj (information about languages). Also, the trials, conditions, interventions, locations and references on LinkedCT have been linked to their related web pages on ClinicalTrials.gov, Wikipedia, Freebase and Pubmed using `foaf:page` links. Table 3 shows the statistics of the external links in LinkedCT.

**Table 3.** Statistics of the links between LinkedCT and several other data sources.

| Source / Target | Link Type | Count |
| --- | --- | --- |
| LinkedCT (intervention) ↔ DBpedia (drug) | owl:sameAs | 11,527 |
| LinkedCT (intervention) ↔ DrugBank (drug) | rdfs:seeAlso | 23,493 |
| LinkedCT (intervention) ↔ DailyMed (drug) | rdfs:seeAlso | 39,396 |
| LinkedCT (condition) ↔ DBpedia (disease) | owl:sameAs | 342 |
| LinkedCT (condition) ↔ Diseasome (disease) | owl:sameAs | 830 |
| LinkedCT (trial) → Geonames | foaf:based_near | 129,177 |
| LinkedCT (reference) → Bio2RDF's PubMed | owl:sameAs | 42,219 |
| LinkedCT (trial) → ClinicalTrials.gov | foaf:page | 61,920 |

The difficulty in discovering links between different data sources lies in the absence of globally unique identifiers in many of the disparate data sources. The problem is exacerbated by presence of errors in the data. The consequence is that establishing links using exact matching of data fields is not robust enough to capture the amount of real linkages. This calls for *approximate string matching* in combination with *semantic matching* of the data fields. Approximate string matching is required when there are different string representations for the same entity. For example, "Alzheimer's disease" in LinkedCT should be matched with "Alzheimer_disease" in Diseasome. String matching is also robust against misspelled words and alternative spellings (eg. "Thalassemia" and "Thalass*a*emia").

Semantic matching is particularly useful for matching clinical terms, since many drugs and diseases have multiple names. Drugs may have generic names and several brand names. Diseases may also have different scientific and non-scientific names or may be abbreviated. For example, "Acquired Immunodeficiency Syndrome" is abbreviated as "AIDS", or the terms "Klinefelter Syndrome" and "Hypogonadotropic hypogonadism" are alternatively used to refer to the same disease. Obviously, approximate string matching is not sufficient for data with such different representations.

**String Matching for Semantic Link Discovery**

String data is prone to several types of inconsistencies and errors including typos, spelling mistakes, using abbreviations or different conventions. Therefore, finding similar strings, or approximate string matching, is an important part of an (online) link discovery process. Approximate string matching is performed based on a similarity function `sim()` that quantifies the amount of closeness (as opposed to distance) between two strings. A similarity threshold $\theta$ is set by the user to specify that there is a link from the base record to the target record if their similarity score, returned by function `sim()`, is above $\theta$.

**Similarity Function** There exists a variety of similarity functions for string data in the literature. The performance of a similarity functions usually depends on the characteristics of data, such as length of the strings, along with the type of the errors and inconsistencies present in the data. An increasingly popular class of string similarity functions is based on tokenization of the strings into q-grams, i.e., substrings of length q of the strings. Using q-gram tokens makes it possible to treat strings as sets of tokens and use a set similarity measure as the measure of similarity between the two strings. Based on comparison of several such measures, we use weighted Jaccard similarity along with q-gram tokens as the measure of our choice due to its relatively high efficiency and accuracy compared with other measures, especially in matching records in medical domain. In particular, this similarity function has the ability to take into account the importance of the tokens, and therefore could match two records that differ in many *unimportant* tokens. This allows matching "HIV virus" with "HIV", and "Alzheimer" with "Alzheimer's Disease". Moreover, this function captures token swaps, therefore matching "Adenocarcinoma of the Colon" with "Colon adenocarcinoma" for example. Many traditional string similarity functions (such as edit similarity) fail to capture such differences in the strings.

Jaccard similarity between two string records $r_1$ and $r_2$ is the fraction of tokens in $r_1$ and $r_2$ that are present

---
[‡‡] http://bio2rdf.org/

in both. Weighted Jaccard similarity is the weighted version of Jaccard similarity, i.e.,

$$sim_{WJaccard}(r_1, r_2) = \frac{\sum_{t \in \mathbf{r}_1 \cap \mathbf{r}_2} w(t,R)}{\sum_{t \in \mathbf{r}_1 \cup \mathbf{r}_2} w(t,R)}$$

where *w(t,R)* is a weight function that reflects the commonality of the token *t* in the relation *R*, i.e., the higher the weight of a q-gram, the more important the q-gram is. We choose RSJ (Robertson/Sparck Jones) weight for the tokens which are similar to (but more effective than) the commonly-used Inverse Document Frequency (IDF) weights:

$$w(t,R) = log\left(\frac{N - n_t + 0.5}{n_t + 0.5}\right)$$

where *N* is the number of tuples in the base relation *R* and $n_t$ is the number of tuples in *R* containing the token *t*. The similarity value returned is between 0 (for strings that do not share any q-grams) and 1 (for equal strings).

**Implementation** We have implemented the weighted Jaccard similarity predicate declaratively in standard SQL and used it as a join predicate in a standard RDBMS engine by extending our previous work on approximate selection[7]. Alternatively, this predicate can be used with some of the specialized, high performance, state-of-the-art approximate join algorithms[4] that make the process scalable to very large databases. In our experiments we used q-grams of size 2 and a specific q-gram generation technique to enhance the effectiveness of the predicate: strings are first padded with whitespaces at the beginning and the end, then all whitespace characters are replaced with q – 1 special symbols (e.g., $).

**Ontology-based Semantic Matching for Link Discovery**

In order to find meaningful semantic links between values, it is often necessary to leverage some semantic knowledge about the domain from which these values originate. In many domains, including health care, there are existing, commonly accepted, semantic knowledge bases that can be used to this end. A common type of such semantic knowledge is an ontology. In the health care domain, well-known ontologies such as the NCI thesaurus are widely used that encapsulate a number of diverse relationship types between their recorded medical terms, including, synonymy, hyponymy/hypernymy, etc. Such relationship types can be conveniently represented in the relational model and (recursive) SQL queries can be used to test whether two values are associated with a relationship of a certain type.

**Implementation** We extend the existing methods for ontology-based keyword search over relational data for ontology-based link discovery. Assume that the semantic knowledge is stored in a table `thesaurus` with columns `src` and `tgt`. The column `src` contains concept IDs of the terms, and the column `tgt` contains the terms. This is a common approach in storing semantic knowledge, used in sources such as NCI thesaurus and Wordnet's synsets. For example, the terms "Acetaminophen", "Tylenol", "Paracetamol" and "APA" all have the same concept ID "Acetaminophen" in the NCI thesaurus. In practice, this data could be stored in a table thesaurus with an additional column *rel* that stores the type of the relationship, or it could be stored in XML format which is a common approach particularly in the health care domain. For the data stored in XML, `thesaurus` can be a view defined in a hybrid XML relational DBMS such as DB2. Due to space constraints, we limit our discussion in this paper to semantic knowledge stored as relational data.

Assume that the trial intervention names are stored in the table `ctintervention(tid,name)` and the DBpedia drug titles are stored in the table `dbpediadrug(tid,title)`. The following SQL query returns names and titles related through the semantic knowledge stored in the table `thesaurus`:

```
SELECT DISTINCT c.*, d.*
FROM ctintervention AS c,
     dbpediadrug AS d, thesaurus AS t
WHERE src IN
      (SELECT src FROM thesaurus
       WHERE tgt = c.name)
      AND tgt = d.title
```

For more complex relationships such as hyponymy or hypernymy, a recursive query can be used to traverse the ontology tree to a specific depth. In our experiments in this paper, we only use synonym information from NCI thesaurus.

**Experiments**

We show the effectiveness of semantic link discovery using string and semantic matching techniques described above using five linkage scenarios in LinkedCT. Table 4 shows the summary of the results. The columns "Link #" show the number of links discovered using exact matching, string matching, semantic matching and the overall number of links. Matching each entity in LinkedCT with the target data source may result in several links. For example, if there are 10 trials on condition "AIDS", and "AIDS" is matched with its corresponding resource on DBpedia, then 10 links are added to the total number of links. The columns "Linked Entity #" show the number of distinct entities in LinkedCT that are linked. The "Diff %" columns show the improvement over the exact matching of the strings. Note that since the links are automatically discovered, some may be inaccurate. These can best be filtered by a user from the query results. The benefit of using automatically discovered links is that a very large number of links is achieved at the cost of some precision. In our case, manually inspecting a subset of the links showed above 98% precision on average in all the scenarios.

**Table 4.** Statistics of the Links Discovered Using String and Semantic Matching.

| | Exact Match | | String Matching | | | | Semantic Matching | | | | Overall | | | |
|---|---|---|---|---|---|---|---|---|---|---|---|---|---|---|
| | Link # | Linked Entity # | Link # | Diff % | Linked Entity # | Diff % | Link # | Diff % | Linked Entity # | Diff % | Link # | Diff % | Linked Entity # | Diff % |
| Intervention ↔ DBpedia (drug) | 8,442 | 867 | 9,716 | +15.1% | 1,558 | +79.7% | 10,630 | 25.9% | 1,334 | +53.9% | 11,527 | **+36.5%** | 1,913 | **+120.6%** |
| Intervention ↔ DrugBank | 9,867 | 926 | 11,865 | +20.2% | 1,938 | +109.3% | 12,127 | +22.9% | 1,641 | +77.2% | 23,493 | **+138.1%** | 2,477 | **+167.5%** |
| Intervention ↔ DailyMed | 14,257 | 673 | 24,461 | +71.6% | 1,296 | +92.6% | 27,685 | +94.2% | 1,099 | +63.3% | 39,396 | **+176.3%** | 1,688 | **+150.8%** |
| Condition ↔ DBpedia | 164 | 164 | 333 | +103.0% | 330 | +101.2% | 173 | +5.5% | 173 | +5.5% | 342 | **+108.5%** | 342 | **+108.5%** |
| Condition ↔ Diseasome | 232 | 192 | 778 | +235.3% | 575 | +199.5% | 301 | +29.7% | 247 | +28.6% | 830 | **+257.8%** | 615 | **+220.3%** |

The results in Table 4 clearly show the effectiveness of both string and semantic matching in all these linkage scenarios. Using semantic knowledge has resulted in more links between intervention and drugs as compared to conditions and diseases since drugs tend to have more semantic equivalences (brand names and generic names). The use of semantic matching however is useful particularly for matching disease names that are abbreviated. The overall numbers of the links show that although there is some overlap between the links discovered using semantic matching and string matching, there is a significant number of links that are not found using a single matching technique.

**Conclusion**

In this paper, we briefly discussed the challenges involved in creating the LinkedCT data space, the first open RDF data source of clinical trials. The data is published using the principles of linked data. Our approach can be used for the publication of other medical datasets as well. Currently, when clinical researchers and pharmaceutical companies bring copies of data within their organizations for integration, they each need to have experts who understand the connectivity across data sets. However, with the Linked Data approach, this responsibility is shifted to the data providers. This is a much more efficient approach, as the data providers are the individuals who understand their data best. It also means that the integration only has to happen one time.

A distinctive feature of LinkedCT data space is the large number of links to other existing data sources. We demonstrated how state-of-the-art approximate string matching and ontology-based semantic matching can be used for discovery of such semantic links between LinkedCT and several other data sources. The ratio of external links per entities in LinkedCT is more than two orders of magnitude larger than the average ratio in the existing linked data sources (that are part of the Linking Open Data community project at W3C). This shows the effectiveness of our proposed techniques in semantic interlinking of trials data and web data sources in general. Due to the declarative nature of our implementation methods, our approach is also useful for matching electronic medical records (EMRs) with trials data. We therefore plan to develop a flexible declarative framework for semantic link discovery over medical databases.

**References**

1. Barrett J, What's behind clinical trial data?(Global News). Journal of Applied Clinical Trials 18.1 (Jan 2009): 22(1).
2. Knowledge@Wharton, Eli Lilly's Sidney Taurel: 'Tailored Therapeutics' -- the Pharmaceutical Industry's Next Blockbuster? March 13, 2008 http://knowledge.wharton.upenn.edu/article.cfm?articleid=1915
3. Berners-Lee T, Linked Data - Design Issues, 07-2006, http://www.w3.org/DesignIssues/LinkedData.html
4. Elmagarmid AK, Ipeirotis PG, Verykios VS, Duplicate record detection: A survey. IEEE Transactions on Knowledge and Data Engineering 2007;19(1), 1–16.
5. Mi, M, Clinical trials database: linking patients to medical research <http//clinicaltrials.gov>. *Journal of Consumer Health on the Internet*, 2005; 9(3): 59-67.
6. Bizer C, Seaborne A, D2RQ - Treating non-RDF databases as virtual RDF graphs, Proceedings of the 3rd International Semantic Web Conference (posters), 2004.
7. Hassanzadeh O, Benchmarking declarative approximate selection predicates. Master's thesis, University of Toronto, Feb. 2007.